\documentclass[twocolumn,aps,showpacs,superscriptaddress,tightenlines,a4paper]{revtex4}

\usepackage{amssymb}
\usepackage{amsmath}
\usepackage{bbold}
\usepackage{mathrsfs}
\usepackage{paralist}
\usepackage{graphicx}
\usepackage{xcolor}

\newcommand{\bc}[1]{\textcolor[rgb]{0.00,0.00,0.60}{\textbf{[#1     $\looparrowright$]}}}

\begin{document}
%
%
	\title{
	Radiation pressure on a moving body: beyond the Doppler effect
	}
%
%
	\author{S. A. R. Horsley}
	\affiliation{School of Physics and Astronomy, University of St Andrews,
North Haugh, St Andrews, KY16 9SS, UK.}
	\email{sarh@st-andrews.ac.uk}
	\affiliation{European Laboratory for Nonlinear Spectroscopy, Sesto
Fiorentino, Italy.}
	\author{M. Artoni}
	\affiliation{European Laboratory for Nonlinear Spectroscopy, Sesto
Fiorentino, Italy.}	
	\affiliation{Department of Physics and Chemistry of Materials CNR-IDASC
Sensor Lab, Brescia University, Brescia, Italy.}
	\author{G. C. La Rocca}
	\affiliation{Scuola Normale Superiore and CNISM, Pisa, Italy.}
%
%
	\begin{abstract}
		The dependence of macroscopic radiation pressure on the velocity of the object being pushed is commonly attributed to the Doppler effect.  This need not be the case, and here we highlight velocity dependent radiation pressure terms that have their origins in the mixing of \emph{s} and \emph{p} polarizations brought about by the Lorentz transformation between the lab and the material rest frame, rather than in the corresponding transformation of frequency and wavevector.  The theory we develop is relevant to the nano-optomechanics of moving bodies.
	\end{abstract}
%
%
	\maketitle
	\bibliographystyle{unsrt}
	\par
	The force of radiation pressure, which arises from the momentum of light, depends upon the velocity of the body being pushed.  As an example, take light incident onto the surface of a perfectly reflecting mirror when this mirror is in motion in the direction of the surface normal.  In the rest frame, the momentum and rate of arrival of the incident photons will appear either increased or decreased relative to the lab, depending whether the mirror moves towards, or away from the light source.  This is a consequence of the Doppler effect, and implies that the velocity dependent part of the radiation pressure acts to slow the mirror down.  When attached to an oscillatory degree of freedom this effect appears as a friction--like term in the equation of motion and is therefore termed \emph{radiation damping}~\cite{braginski1967,matsko1996}.  If the ideal mirror is replaced by a material that is highly dispersive (e.g. a Bragg mirror), the reflection coefficient of the moving medium is also strongly dependent upon the Doppler shifted frequency, and this can turn damping into heating~\cite{horsley2011,favero2008}.  
	\par
	Another instance of a velocity dependent radiation pressure effect is \emph{universal drag} whereby all media in relative motion at finite temperature experience a frictional force due to the electromagnetic field~\cite{mkrtchian2003};  this too has its origins in the Doppler effect.  One consequence of universal drag is that when two polarizable bodies with equal rest frame temperature ($T\neq 0$) move relative to one another, they act to slow one another down via the electromagnetic field.
	\par
	The relative motion of a dielectric medium, however, has a peculiar effect on the polarization of light as well as its propagation. The moving medium behaves as a magnetoelectric~\cite{volume8,leonhardt2010b}, which implies a change in the polarization composition (\(s\) \& \(p\)) upon interacting with the medium.
Here we raise the question whether there exist optical forces, at least from a classical standpoint~\footnote{There has been some debate surrounding the role of the non--Doppler part of the force due to the electromagnetic vacuum quantum fluctuations for moving plates (Casimir friction).  On one side it has been argued that there should be a frictional force between the plates~\cite{pendry1997}, but this calculation ignores non--Doppler contributions to the force.  Another calculation, which takes these into account~\cite{philbin2009} has claimed a zero value for the frictional force.}, arising from polarization mixing effects in a moving medium, in spite of the fact that many features of reflection from, and transmission through moving dielectrics have been considered before~\cite{yeh1966,huang1994}.  What's more, is there an interplay between polarization mixing effects and Doppler effect that is important to the radiation pressure experienced by a moving dielectric?  Here we show that this is indeed the case, and that there are situations where polarization mixing effects occur at the same order of \(V/c\) as the Doppler effect, and hence cannot be neglected.  Our theoretical framework, based on generalized scattering matrix methods, is readily amenable to single out these non-Doppler contributions to the radiation pressure experienced by a moving dielectric. We further examine a clear-cut configuration where the effect of radiation pressure is solely due to polarization mixing.
%
\begin{figure}
	\includegraphics[width=6cm]{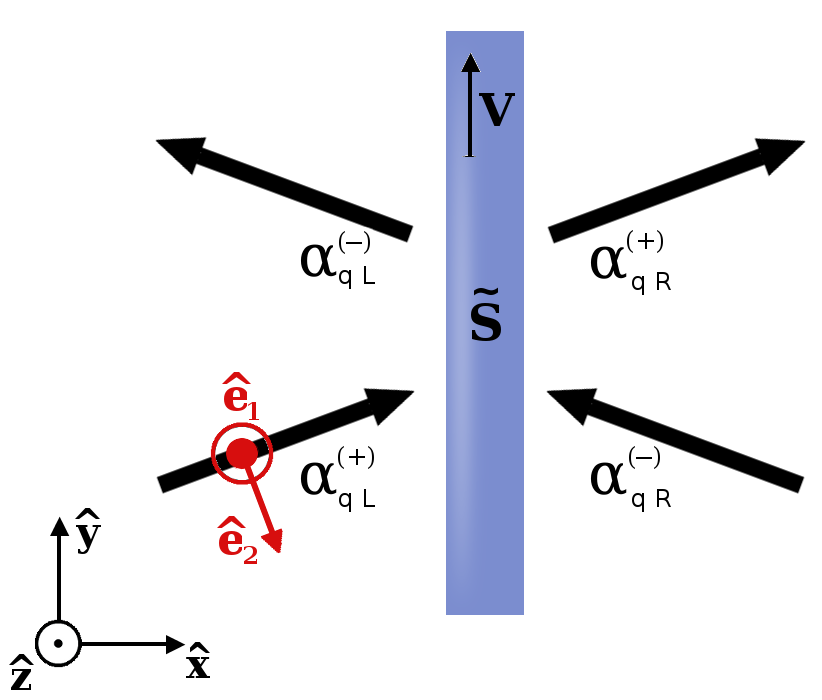}
	\caption{In the lab frame the \(\boldsymbol{\tilde{S}}\) matrix connects input ($\alpha_{\text{\tiny{q\,L}}}^{\text{\tiny{($+$)}}},\alpha_{\text{\tiny{q\,R}}}^{\text{\tiny{($-$)}}}$) and output ($\alpha_{\text{\tiny{q\,L}}}^{\text{\tiny{($-$)}}},\alpha_{\text{\tiny{q\,R}}}^{\text{\tiny{($+$)}}}$)  amplitudes of fields scattering from the thin slab moving laterally with velocity  \(\boldsymbol{V}\). The $s$ and $p$ polarization unit vectors, respectively, $\hat{\boldsymbol{e}}_{1}$ and $\hat{\boldsymbol{e}}_{2}$ are here displayed for the special case
	 of a plane wave propagating  in the (\(x-y\)) plane with $\hat{\boldsymbol{k}}_{\parallel} || \  \hat{\boldsymbol{y}}$.
	   More generally the unit vectors $\hat{\boldsymbol{e}}_{1}$ and $\hat{\boldsymbol{e}}_{2}^{\text{\tiny{($+$)}}}$ lie in the ($\hat{\boldsymbol{z}}-\hat{\boldsymbol{y}}$) plane and in the incidence plane, i.e. the ($\hat{\boldsymbol{x}}-\hat{\boldsymbol{k}}_{\parallel}$) plane, respectively.
	 \label{figure-1}}
	\end{figure}
\par
	The lab frame electric field of radiation  exerting pressure on a moving dielectric slab (see figure~\ref{figure-1}) can  be written as,
	\begin{equation}
	  \boldsymbol{E}(\boldsymbol{x},t)=\sum_{\pm,q=1,2}\hat{\boldsymbol{e}}_{q}^{\text{\tiny{($\pm$)}}}\alpha^{\text{\tiny{($\pm$)}}}_{q}e^{i(\pm|k_{x}|x+\boldsymbol{k}_{\parallel}\cdot\boldsymbol{x}-\omega t)}.\label{electric-field-lab}
	\end{equation}
where the surface has a normal parallel to \(\hat{\boldsymbol{x}}\), 
 \(\boldsymbol{k}_{\parallel}=k_{y}\hat{\boldsymbol{y}}+k_{z}\hat{\boldsymbol{z}}\) and \(\boldsymbol{k}^{\text{\tiny{($\pm$)}}}=\pm |k_x|\,\hat{x}+\boldsymbol{k}_{\parallel}\), with all wavevector components real and with polarization unit vectors \(\hat{\boldsymbol{e}}_{1}^{\text{\tiny{($\pm$)}}}=\hat{\boldsymbol{e}}_{1}=\hat{\boldsymbol{x}}\boldsymbol{\times}\hat{\boldsymbol{k}}_{\parallel}\), \(\hat{\boldsymbol{e}}_{2}^{\text{\tiny{($\pm$)}}}={\hat{\boldsymbol{k}}}^{\text{\tiny{($\pm$)}}}\boldsymbol{\times}\hat{\boldsymbol{e}}_{1}\),  that are equivalent to \emph{s} (\(q=1\)) and \emph{p} (\(q=2\)) polarization, respectively.  The electric field transformation into the rest (primed) frame of the moving slab depends upon its direction of motion.  In particular, for (\(\boldsymbol{V}=V_{x}\hat{\boldsymbol{x}}\)), the rest frame field amplitudes, \({\alpha^{\prime}}_{l}^{\text{\tiny{($\pm$)}}}\) are found to be related to the lab frame by a constant of proportionality, \({\alpha^{\prime}}_{l}^{\text{\tiny{($\pm$)}}}=({\omega^{\prime}}^{\text{\tiny{($\pm$)}}}/\omega)\alpha_{l}^{\text{\tiny{($\pm$)}}}\), with \({\omega^{\prime}}^{\text{\tiny{($\pm$)}}}=\gamma(\omega\mp V_{x}|k_{x}|)\) and \(\gamma=(1-\boldsymbol{V}^{2}/c^{2})^{-1/2}\).  Meanwhile for \textit{lateral} motion (\(\boldsymbol{V}=V_{y}\hat{\boldsymbol{y}}\)) we find, \(\boldsymbol{\alpha}^{\prime}=(\omega^{\prime}/\omega)\boldsymbol{M}\boldsymbol{\cdot}\boldsymbol{\alpha}\), where,
	\begin{equation}
		\boldsymbol{M}=\frac{1}{\sqrt{1+\frac{\eta^{2}V_{y}^{2}}{c^{2}}}}\left(\begin{matrix}\boldsymbol{\mathbb{1}}_{2}&\frac{V_{y}\eta}{c}\boldsymbol{\sigma}_{z}\\
	                                                                              -\frac{V_{y}\eta}{c}\boldsymbol{\sigma}_{z}&\boldsymbol{\mathbb{1}}_{2}
	                                                                             \end{matrix}\right),\label{lateral-motion}		
	\end{equation}
	\(\omega^{\prime}=\gamma(\omega-V_{y}k_{y})\), \(\eta=|k_{x}|k_{z}/(k_{\parallel}^{2}-V_{y}\omega k_{y}/c^{2})\), \(\boldsymbol{\alpha}^{\text{\tiny{T}}}=\left(\alpha_{1}^{\text{\tiny{(+)}}},\alpha_{1}^{\text{\tiny{(-)}}},\alpha_{2}^{\text{\tiny{(+)}}},\alpha_{2}^{\text{\tiny{(-)}}}\right)\), and \(\boldsymbol{\sigma}_{z}\) is the usual Pauli matrix.   Note that all of the particular velocity dependent radiation pressure effects mentioned in the introduction arise from the Doppler shift, and this holds also here for motion along \(\hat{\boldsymbol{x}}\), where the field amplitudes are only modified by the factors \({\omega^{\prime}}^{\text{\tiny{($\pm$)}}}/\omega\).  
	The situation changes for dielectric media in lateral motion where the amplitudes are modified both by the Doppler shifted frequencies and by the transformation  (\ref{lateral-motion}). The unitary \(4\times4\) matrix  $\boldsymbol{M}$ describes how the two polarizations are coupled or mixed by the motion of the medium, and this specifically hinges on the off diagonal \(2\times2\) matrices within (\ref{lateral-motion}) that depend on the lateral velocity $V_{y}$ and the  parameter $\eta$.  Such a \emph{polarization mixing effect} vanishes for $k_z=0$ in which case the incidence plane contains the $y$ axis along which the slab moves and thus it remains a plane of mirror symmetry for the system assuring that $s$ and $p$ polarizations be independent.  Conversely, when $k_z \ne 0$ and \(k_{\parallel}^{2}=V_{y}\omega k_{y}/c^{2}\), we have \(\eta\to\infty\), and the polarization is now completely converted from one type to another.
\begin{figure}
	\includegraphics[width=7.5cm]{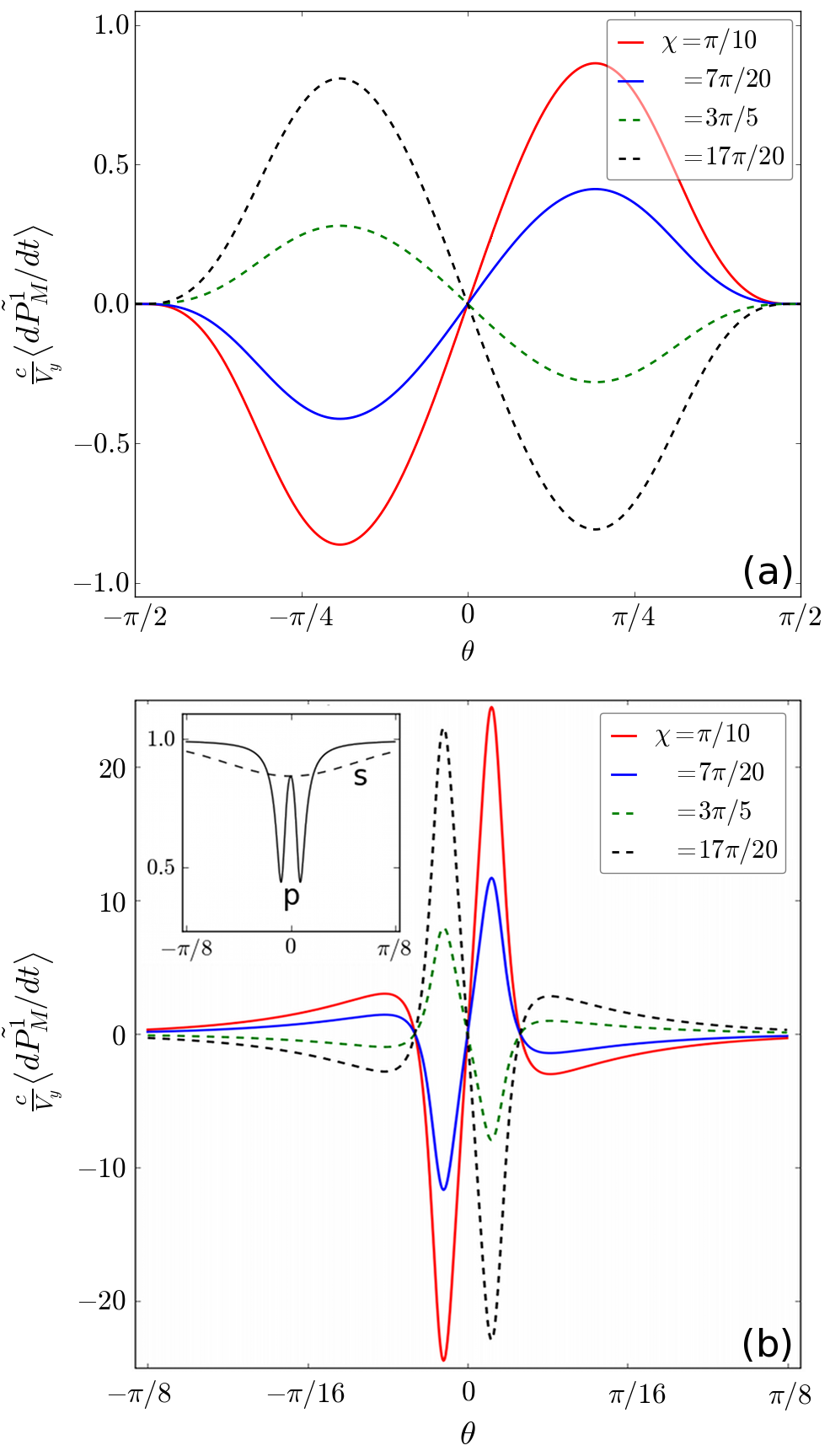}
	\caption{(a) Normal  force (\ref{normal-force-2}) as a function of the angle of incidence $\theta$, acting on a slab of thickness \(5c/\omega\) with (a) \(\epsilon=5.0+0.01i\) and (b)  \(\epsilon=0.001+0.001i\) and \(\mu=1\).  The incident wave--vector is written as, \(\boldsymbol{k} = k_x \hat{\boldsymbol{x}} + k_{||}\hat{\boldsymbol{k}}_{||} = (\omega/c)[\cos(\theta)\hat{\boldsymbol{x}}+\sin(\theta)(\sin(\chi)\hat{\boldsymbol{y}}+\cos(\chi)\hat{\boldsymbol{z}})]\).  Inset shows the \(s\) (dashed) and \(p\) (solid) reflection coefficients for such a slab .\label{figure-2}}
	\end{figure}
We wish to investigate to what extent mixing of the $s$ and $p$ polarization composition affects radiation pressure when light is incident onto a moving medium. We will then consider in what follows only the case of a laterally moving slab of thickness, \(d\), centred at \(x,y,z=0\), and of \(y\)--\(z\) area, \(A=L_{y}L_{z}\), where \(L_{y},L_{z}\gg d\).  In both free space regions, \(x<-d/2\) (L), and \(x>d/2\) (R), we assume an electromagnetic field in the form of a right--going wave plus a left--going wave as in (\ref{electric-field-lab}).  In the case of a slab in lateral motion, the frequency is conserved upon reflection and transmission, and so we consider all waves to have a single frequency, \(\omega\).
		The average four--force experienced by the slab can be written as the difference between input and output photon flux in each polarization, directed along the wave--vector, 
		\(k_\pm^{\mu}=(\omega/c,\pm|k_{x}|,\boldsymbol{k}_{\parallel})\),
	\begin{equation}
	    \left\langle\frac{d P_{\text{\tiny{M}}}^{\mu}}{dt}\right\rangle=\frac{\epsilon_{0}A}{2}\sum_{q,\pm}\left(\frac{c}{\omega}\right)^{2}\left(|\alpha_{q\,\text{\tiny{L}}}^{\text{\tiny{($\pm$)}}}|^{2}-|\alpha_{q\,\text{\tiny{R}}}^{\text{\tiny{($\pm$)}}}|^{2}
	    \right)k_{x}k_\pm^{\mu}.\label{final-four-force}
	\end{equation}
	This expression can be derived from a consideration of the energy--momentum tensor averaged over a time interval, \(T\gg\omega^{-1}\), as in~\cite{horsley2011}.  In general, we have knowledge of only the input fields, \(\boldsymbol{\alpha}_{\text{\tiny{IN}}}^{\text{\tiny{T}}}=(\alpha_{1\,\text{\tiny{L}}}^{\text{\tiny{($+$)}}},\alpha_{1\,\text{\tiny{R}}}^{\text{\tiny{($-$)}}},\alpha_{2\,\text{\tiny{L}}}^{\text{\tiny{($+$)}}},\alpha_{2\,\text{\tiny{R}}}^{\text{\tiny{($-$)}}})\), and not the output fields, \(\boldsymbol{\alpha}_{\text{\tiny{OUT}}}^{\text{\tiny{T}}}=(\alpha_{1\,\text{\tiny{R}}}^{\text{\tiny{($+$)}}},\alpha_{1\,\text{\tiny{L}}}^{\text{\tiny{($-$)}}},\alpha_{2\,\text{\tiny{R}}}^{\text{\tiny{($+$)}}},\alpha_{2\,\text{\tiny{L}}}^{\text{\tiny{($-$)}}})\).  The relationship between the two is given by the scattering matrix, \(\boldsymbol{S}\): \(\boldsymbol{\alpha}_{\text{\tiny{OUT}}}=\boldsymbol{S}\boldsymbol{\cdot}\boldsymbol{\alpha}_{\text{\tiny{IN}}}\).  In our case the motion of the medium will mix the polarizations, and we must deal with a \(4\times4\) scattering matrix, rather than the standard \(2\times2\) object~\cite{genet2003}, the elements of which are written in general as,
	\begin{equation}
	    \boldsymbol{S}=\left(\begin{matrix}\boldsymbol{S}_{11}&\boldsymbol{S}_{12}\\\boldsymbol{S}_{21}&\boldsymbol{S}_{22}\end{matrix}\right)=\left(\begin{matrix}\mathcal{T}_{11}&\bar{\mathcal{R}}_{11}&\mathcal{T}_{12}&\bar{\mathcal{R}}_{12}\\
	                         \mathcal{R}_{11}&\bar{\mathcal{T}}_{11}&\mathcal{R}_{12}&\bar{\mathcal{T}}_{12}\\
				 \mathcal{T}_{21}&\bar{\mathcal{R}}_{21}&\mathcal{T}_{22}&\bar{\mathcal{R}}_{22}\\
				 \mathcal{R}_{21}&\bar{\mathcal{T}}_{21}&\mathcal{R}_{22}&\bar{\mathcal{T}}_{22}
	                         \end{matrix}\right).\label{scattering-matrix}
	\end{equation}
	In the case of media which do not mix \(s\) and \(p\) polarizations, (\ref{scattering-matrix}) reduces to a direct sum, \(\boldsymbol{S}=\boldsymbol{S}_{11}\boldsymbol{\oplus}\boldsymbol{S}_{22}\).
      \par
      Using (\ref{scattering-matrix}), the radiation pressure experienced by the moving slab can be written in terms of the incident field amplitudes and the rest frame scattering matrix, \(\boldsymbol{S}^{\prime}(k_\pm^{\prime\,\mu})\).  First, (\ref{final-four-force}) is written in the rest frame of the medium in terms of \(\boldsymbol{S}^{\prime}\) and the field amplitudes, \(\boldsymbol{\alpha}^{\prime}_{\text{\tiny{IN}}}\).  After performing a Lorentz transformation back to the lab frame, and applying (\ref{lateral-motion}), we find the four--force in the lab frame in terms of lab frame field amplitudes,
      \begin{equation}
	  \left\langle\frac{dP^{\mu}_{\text{\tiny{M}}}}{dt}\right\rangle=\frac{\epsilon_{0}Ac^{2}|k_{x}|}{2\omega^{2}}
	  \boldsymbol{\alpha}_{\text{\tiny{IN}}}^{\dagger}\left(\begin{matrix}\left(\mathbb{1}_{4}-
	  \tilde{\boldsymbol{S}}^{\dagger}\tilde{\boldsymbol{S}}\right)\omega/c&\\
	  (\boldsymbol{R}-\tilde{\boldsymbol{S}}^{\dagger}\boldsymbol{R}\tilde{\boldsymbol{S}})|k_{x}|\\
	  (\mathbb{1}_{4}-\tilde{\boldsymbol{S}}^{\dagger}\tilde{\boldsymbol{S}})k_{y}\\
	  (\mathbb{1}_{4}-\tilde{\boldsymbol{S}}^{\dagger}\tilde{\boldsymbol{S}})k_{z}
	  \end{matrix}\right)\boldsymbol{\alpha}_{\text{\tiny{IN}}},\label{final-result}
      \end{equation}
      where, \(\boldsymbol{R}=\boldsymbol{\sigma}_{z}\boldsymbol{\oplus}\boldsymbol{\sigma}_{z}\), and the field amplitudes outside of the vector multiply the matrices inside.  The lab frame scattering matrix, \(\tilde{\boldsymbol{S}}\) is related to the rest frame matrix by, \(\tilde{\boldsymbol{S}}=\boldsymbol{M}^{\dagger}\boldsymbol{S}^{\prime}\boldsymbol{M}\).  The conclusion we can draw from comparing (\ref{final-result}) with its limit when \(V_{y}\to0\) is that a laterally moving medium responds akin to a stationary medium with a peculiar frequency response, and reflection and transmission coefficients that are related to the rest frame reflection and transmission coefficients by a unitary transformation. At a frequency $\omega^\prime$ where the material has negligible loss, \(\boldsymbol{S}^{\prime\dagger}\boldsymbol{S}^{\prime}=\mathbb{1}_{4}\), all components of the four-force, besides \(\left\langle dP^{1}_{\text{\tiny{M}}}/dt\right\rangle\), vanish.  Polarizations are not mixed in the rest frame where one has \(\boldsymbol{S^\prime}=\boldsymbol{S^\prime}_{11}\boldsymbol{\oplus}\boldsymbol{S^\prime}_{22}\) with \(\mathcal{R}_{ii}^{\prime}=\bar{\mathcal{R}}^{\prime}_{ii}=r_{i}^{\prime}\), \& \(\mathcal{T}_{ii}^{\prime}=\bar{\mathcal{T}}^{\prime}_{ii}=t_{i}^{\prime}\), where $r_{1,2}^{\prime}$ and $t_{1,2}^{\prime}$ are respectively the rest frame reflection and transmission amplitudes for $s$ and $p$ polarization.  In this case one has the lab frame scattering matrix,
      \begin{widetext}
	    \begin{equation}\tilde{\boldsymbol{S}}=\boldsymbol{M}^{\dagger}\left(\boldsymbol{S}^{\prime}_{11}\boldsymbol{\oplus}\boldsymbol{S}_{22}^{\prime}\right)\boldsymbol{M}=\frac{1}{1+\frac{\eta^{2}V_{y}^{2}}{c^{2}}}\left(\begin{matrix}\boldsymbol{S}_{11}^{\prime}+\left(\frac{\eta V_{y}}{c}\right)^{2}\boldsymbol{\sigma}_{z}\boldsymbol{S}_{22}^{\prime}\boldsymbol{\sigma}_{z}&
	      \frac{\eta V_{y}}{c}\left(\boldsymbol{S}^{\prime}_{11}\boldsymbol{\sigma}_{z}-\boldsymbol{\sigma}_{z}\boldsymbol{S}^{\prime}_{22}\right)\\
	      \frac{\eta V_{y}}{c}\left(\boldsymbol{\sigma}_{z}\boldsymbol{S}^{\prime}_{11}-\boldsymbol{S}^{\prime}_{22}\boldsymbol{\sigma}_{z}\right)&
	      \boldsymbol{S}^{\prime}_{22}+\left(\frac{\eta V_{y}}{c}\right)^{2}\boldsymbol{\sigma}_{z}\boldsymbol{S}^{\prime}_{11}\boldsymbol{\sigma}_{z}\end{matrix}\right)\label{transformed-S-matrix}
	    \end{equation}
      \end{widetext}
where the off-diagonal blocks linear in $(\eta V_y/c)$ are responsible for the polarization mixing effects.

The four--force in (\ref{final-result}) along with (\ref{transformed-S-matrix}) is an important result, as it embeds a new radiation pressure effect stemming from mixing of the $s$ and $p$ polarizations associated with moving frames. Two specific examples are now illustrated. First, take the case of a plane polarized wave propagating in the direction  $\hat{\boldsymbol{k}}^{\text{\tiny{($+$)}}}$ onto the left of a laterally moving slab, the two input polarization 
amplitudes being  $\alpha_{1\,\text{\tiny{L}}}^{\text{\tiny{(+)}}}$ and $\alpha_{2\,\text{\tiny{L}}}^{\text{\tiny{(+)}}}$.  The resulting normalized~\footnote{The tilde indicates that the force is in units of the incident average power \(\epsilon_{0} c (|\alpha_{1\,\text{\tiny{L}}}^{\text{\tiny{(+)}}}|^2+|\alpha_{2\,\text{\tiny{L}}}^{\text{\tiny{(+)}}}|^2)\,A/2\) divided by $c$.} normal (\(\hat{\boldsymbol{x}}\)) force  is,
\begin{widetext}
\begin{multline}
	\tilde{\left\langle\frac{dP^{1}_{\text{\tiny{M}}}}{dt}\right\rangle}=
\frac{c^{2}k_{x}^{2}}{\omega^{2}}\left(1+a_{1}\,\left|\alpha_1+\frac{\eta V_{y}}{c}\alpha_2\right|^{2}
		+a_{2}\,\left|\alpha_2-\frac{\eta V_{y}}{c}\alpha_1\right|^{2}\right)\\
\simeq \frac{c^{2}k_{x}^{2}}{\omega^{2}}\; \Big[ (2R^\prime_1+A^\prime_1)|\alpha_1|^2+(2R^\prime_2+A^\prime_2)|\alpha_2|^2 +
\frac{|k_x|k_z}{k^2_\parallel}\frac{V_y}{c}(2R_1+A_1-2R_2-A_2)(\alpha_1^*\alpha_2 +c.c.) \Big]
\label{normal-force}
	\end{multline}
\end{widetext}
	where we  define   $\alpha_{i}=\alpha_{i\,\text{\tiny{L}}}^{\text{\tiny{(+)}}}/{\sqrt{|\alpha_{1\,\text{\tiny{L}}}^{\text{\tiny{(+)}}}|^2+|\alpha_{2\,\text{\tiny{L}}}^{\text{\tiny{(+)}}}|^2}}$,  \(a_{i}=(\left|r^{\prime}_{i}\right|^{2}-\left|t^{\prime}_{i}\right|^{2})/(1+\eta^{2}V_{y}^{2}/c^{2})=(2R^\prime_i+A^\prime_i-1)/(1+\eta^{2}V_{y}^{2}/c^{2})\).
	The last expression is correct to first order in $V_y/c$ ($\eta\simeq|k_x|k_z/k^2_\parallel$).   
Other terms linear in $V_y/c$ would appear upon expressing $R^\prime_i$ and $A^\prime_i$ in terms of the lab frame quantities through the transformation of frequency and wavector components. Such an expansion would bring out familiar Doppler terms that are independent of the polarization mixing terms proportional to $(\alpha_1^*\alpha_2 +c.c.)$   shown on the right hand side of (\ref{normal-force}).

The normalized lateral (\(\hat{\boldsymbol{y}}\)) force, which vanishes for a lossless medium ($A_i^\prime=0$), takes a similar form, 
\begin{widetext}
	\begin{multline}
	\tilde{\left\langle\frac{dP^{2}_{\text{\tiny{M}}}}{dt}\right\rangle}=
\frac{c^{2}|k_{x}| k_y}{\omega^{2}}\left(1+b_{1}\,\left|\alpha_1+\frac{\eta V_{y}}{c}\alpha_2\right|^{2}
		+b_{2}\,\left|\alpha_2-\frac{\eta V_{y}}{c}\alpha_1\right|^{2}\right)\\
\simeq \frac{c^{2}|k_{x}| k_y}{\omega^{2}} \; \Big[A^\prime_1|\alpha_1|^2+A^\prime_2|\alpha_2|^2
+\frac{|k_x|k_z}{k^2_\parallel}\frac{V_y}{c}(A_1-A_2)(\alpha_1^*\alpha_2 +c.c.) \Big]
\label{lateral-force}
	\end{multline}
\end{widetext}
with \(b_{i}=(-\left|r^{\prime}_{i}\right|^{2}-\left|t^{\prime}_{i}\right|^{2})/(1+\eta^{2}V_{y}^{2}/c^{2})=(A^\prime_i-1)/(1+\eta^{2}V_{y}^{2}/c^{2})\).  In general, polarization mixing contributions to the velocity dependent part of the force (\ref{normal-force}-\ref{lateral-force}) are comparable to the Doppler terms \footnote{This holds for weakly dispersive media: whenever $R^\prime$ and $A^\prime$ have a strong dependence on frequency, such as in photonic bandgap media or close to an absorption line, the Doppler related effects can be amplified by many orders of magnitude~\cite{favero2008,horsley2011}}.  Since they are proportional to $(\alpha_1^*\alpha_2 +c.c.)$, they vanish for unpolarized light, such as {\it e.g.} thermal radiation, where $\alpha_1$ and $\alpha_2$ do not have a definite mutual phase relation, and also for circularly polarized light ($\alpha_1=1/\sqrt{2}$, $\alpha_2=\pm i/\sqrt{2}$).  Even when $\eta\ne 0$, the polarization mixing effects vanish to linear order in $V_y/c$ if in the lab frame the light is either  purely $s$ or purely $p$ polarized, yet they survive and could be easily modulated or even flipped in sign by rotating the linear polarization when reflection and transmission for $s$ and $p$ polarizations are significantly different. In particular, for linearly polarized light with $\alpha_1=\sin\varphi$ and $\alpha_2=\cos\varphi$, corresponding to an  electric field whose polarization vector makes an angle $\varphi$ with $\hat{\boldsymbol{e}}_{2}^{\text{\tiny{($+$)}}}$, one has $(\alpha_1^*\alpha_2 +c.c.)=\sin(2\varphi)$.
	\par
	As a second example, we consider a situation where radiation pressure arises only when both the mixing coefficient \(\eta\) and the velocity do not vanish.  Take two linearly polarized plane waves of equal intensity propagating in the direction $\hat{\boldsymbol{k}}^{\text{\tiny{($+$)}}}$ and $\hat{\boldsymbol{k}}^{\text{\tiny{($-$)}}}$, respectively onto the left and the right of a laterally moving slab.  In much the same way as done above, the input field amplitudes may be chosen so that \(\boldsymbol{\alpha}_{\text{\tiny{IN}}}^{\text{\tiny{T}}}={\cal E}(e^{i\zeta_{\text{\tiny{L}}}}\sin(\varphi_L),e^{i\zeta_{\text{\tiny{R}}}}\sin(\varphi_R),e^{i\zeta_{\text{\tiny{L}}}}\cos(\varphi_L),e^{i\zeta_{\text{\tiny{R}}}}\cos(\varphi_R))\), where 
\(\Delta\zeta=\zeta_{\text{\tiny{L}}}-\zeta_{\text{\tiny{R}}}\) is the phase mismatch between the two input waves. Note that with our choice of polarizations the plane wave coming from the right and having $\varphi_R=\pi-\varphi_L$ is the mirror image of that coming from the left with respect to the (\(z\)-\(y\)) plane of symmetry. Taking an average over $\Delta\zeta$ as appropriate for left and right incoming waves not having a definite mutual phase relation, the normal component of the force in this case can be obtained from (\ref{normal-force}) and the use of the mirror symmetry just described. In particular, for the specific choice $\varphi_R=\varphi_L=\varphi$ and retaining only terms linear in \(V_{y}/c\), one has
\begin{align}
		\tilde{\left\langle\frac{dP^{1}_{\text{\tiny{M}}}}{dt}\right\rangle}&\simeq\frac{2c^{2}k_{x}^{2}}{\omega^{2}}\frac{|k_x|k_z}{k^2_\parallel}\frac{V_y}{c}[2(R_1-R_2)+A_1-A_2]\sin(2\varphi)\label{normal-force-2}
\end{align}  
	\par
	This force, that is present even when equal incident intensities are used (in both polarization channels) on both sides of the slab, is entirely due to the asymmetry in the polarization transformation between left and right propagating waves as described by (\ref{lateral-motion}) and does not seem to have been recognised before.  In addition, the force (\ref{normal-force-2}), which does not contain Doppler contributions, is typically of the same order as the velocity dependent radiation pressure stemming from the Doppler effect as mentioned in the introduction. Figure \ref{figure-2}a shows the dependence of this force on the angle of incidence for a thin dielectric slab when $\varphi=\pi/4$.  Again, there are clearly situations that make (\ref{normal-force-2}) vanish altogether, such as when we have equal reflection and absorption for the two polarizations, {\it e.g.} for surfaces behaving as ideal mirrors.  
	\par
	For small (but not zero) \(k_{\parallel}\), as \(\eta\sim|k_{x}|k_{z}/k_{\parallel}^{2}\) becomes very large, there is a large degree of polarization mixing.  For ordinary materials this does not have an effect due to the near degeneracy of \(s\) and \(p\) reflection and transmission coefficients at small angles of incidence.  However, there are materials where the reflection and transmission coefficients remain significantly different close to normal incidence.  For example, such a dependence on angle could apply to epsilon near zero metamaterials~\cite{alu2007,alekseyev2010}, and figure \ref{figure-2}b shows more than an order of magnitude increase in the polarization mixing contribution to the velocity dependent force.
	\par
	The configuration leading to (\ref{normal-force-2}) is of special interest as it enables one to observe \textit{pure} polarization mixing contributions to the velocity dependence of radiation pressure. This, in turn, can simply be tuned via the polarization angle, for instance it can be turned on and off by setting $\varphi_L=\pi/4$ and switching $\varphi_R$ respectively to $\pi/4$ and $3\pi/4$. Such an ability can be important in  the emerging field of nano-optomechanics \cite{marquardt2009}, where small optical forces can be used to manipulate the centre of mass of mechanical oscillators of very light masses, or even atomic clouds~\cite{horsley2011}.  In particular, it can allow to control the dynamics of a two-dimensional oscillator in which the motion along one axis ($\hat{\boldsymbol{y}}$ in our case) can be optomechanically coupled to that along the other one ($\hat{\boldsymbol{x}}$ in our case). 
        \acknowledgments
        One of us (SARH) would like to thank the EPSRC for financial support.  This work was also supported by the CRUI-British Council Programs ``Atoms and Nanostructures" and ``Metamaterials", and the IT09L244H5 Azione Integrata MIUR grant.
        \bibliography{pmrp-refs}

\begin{thebibliography}{10}

\bibitem{braginski1967}
V.~B. Braginski and A.~B. Makunin.
\newblock {\em J. Exp. Teor. Phys.}, 52:998, 1967.

\bibitem{matsko1996}
A.~B. Matsko, E.~A. Zubova, and S.~P. Vyatchanin.
\newblock {\em Opt. Commun.}, 131:107, 1996.

\bibitem{horsley2011}
S.~A.~R. Horsley, M.~Artoni, and G.~C. La~Rocca.
\newblock {\em Phys. Rev. Lett.}, 107:043602, 2011.

\bibitem{favero2008}
K.~Karrai, I.~Favero, and C.~Metzger.
\newblock {\em Phys. Rev. Lett.}, 100:240801, 2008.

\bibitem{mkrtchian2003}
V.~Mkrtchian, V.~Parsegian, R.~Podgornik, and W.~M. Saslow.
\newblock {\em Phys. Rev. Lett.}, 91:220801--1, 2003.

\bibitem{volume8}
L.~D. Landau, E.~M. Lifshitz, and L.~P. Pitaevskii.
\newblock {\em The Electrodynamics of Continuous Media}.
\newblock Butterworth-Heinemann, Oxford, 2004.

\bibitem{leonhardt2010b}
U.~Leonhardt and T.~G. Philbin.
\newblock {\em Geometry and Light: The Science of Invisibility}.
\newblock Dover, New York, 2010.

\bibitem{yeh1966}
C.~Yeh.
\newblock {\em J. Appl. Phys.}, 37:3079, 1966.

\bibitem{huang1994}
Yao-Xiong Huang.
\newblock {\em J. Appl. Phys.}, 76:2575, 1994.

\bibitem{genet2003}
C.~Genet, A.~Lambrecht, and S.~Reynaud.
\newblock {\em Phys. Rev. A}, 67:043811, 2003.

\bibitem{alu2007}
A.~Al{\`u}, M.~G. Silveirinha, A.~Salandrino, and N.~Engheta.
\newblock {\em Phys. Rev. B}, 75:155410, 2007.

\bibitem{alekseyev2010}
L.~V. Alekseyev, E.~E. Narimanov, T.~Tumkur, Yu.~A. Barnakov, and M.~A.
  Noginov.
\newblock {\em Appl. Phys. Lett.}, 97:131107, 2010.

\bibitem{marquardt2009}
F.~Marquardt and S.~M. Girvin.
\newblock {\em Physics}, 2:40, 2009.

\bibitem{pendry1997}
J.~B. Pendry.
\newblock {\em J. Phys. Cond. Mat.}, 9:10301, 1997.

\bibitem{philbin2009}
T.~G. Philbin and U.~Leonhardt.
\newblock {\em New J. Phys.}, 11:033035, 2009.

\end{thebibliography}

\end{document}